\documentclass[final,5p,times,twocolumn]{elsarticle}
\usepackage{lineno,hyperref}
\usepackage{float}
\usepackage{subfig}
\journal{Nuclear Instruments and Methods in Physics Research, Section A, }
\usepackage{eso-pic}
\usepackage{ragged2e}

\begin{document}

\begin{frontmatter}

\title{LUXE-PROC-2022-005\raggedleft\\Detector Challenges of the strong-field QED experiment LUXE at the European XFEL\tnoteref{n1}}

\tnotetext[n1]{on behalf of \textbf{LUXE} and \textbf{FCAL} Collaborations}

\author[iss]{Veta Ghenescu\corref{cor1}\centering}
\cortext[cor1]{Corresponding author}
\ead{veta.ghenescu@cern.ch}
\address[iss]{Institute of Space Science, Atomistilor 409, P.O. Box MG-23, Bucharest-Magurele RO-077125, ROMANIA}


\begin{abstract}
The LUXE experiment aims at studying high-field QED in electron-laser and photon-laser interactions, with the 16.5 GeV electron beam of the European XFEL and a laser beam with power of up to 350 TW. The experiment will measure the spectra of electrons, positrons and photons in the expected range of ${10}^{-3}$
 to ${10}^9$
  per 1 Hz bunch crossing, depending on the laser power and focus. These measurements have to be performed in the presence of low-energy high radiation-background. To meet these challenges, for high-rate electron and photon fluxes, the experiment will use Cherenkov radiation detectors, scintillator screens, sapphire sensors as well as lead-glass monitors for backscattering off the beam-dump. A four-layer silicon-pixel tracker and a compact electromagnetic tungsten calorimeter with GaAs sensors will be used to measure the positron spectra. The layout of the experiment and the expected performance under the harsh radiation conditions, together with the test of the Cherenkov detector and the electromagnetic (EM) calorimeter performed recently at DESY,  are presented. The experiment received a stage 0 critical approval (CD0) from the DESY management and is in the process of preparing its technical design report (TDR). It is expected to start running in 2025/2026.

\end{abstract}


\end{frontmatter}


\section{Introduction}
Quantum Electrodynamics (QED) is an important part of the Standard Model that forms a cornerstone of modern physics.  LUXE (Laser Und XFEL Experiment) is a new experiment proposed at DESY and the European XFEL to study QED in a new, previously unexplored regime of strong electromagnetic background fields where it becomes non-perturbative. To probe this new regime, LUXE will perform collisions between the high energetic electron beam from the European XFEL facility and an optical laser, as well as collisions between a secondary photon beam in the GeV energy regime and the high-power laser. Due to the high laser intensity and the Lorentz-boost of the initial electron beam, LUXE achieves strong field close to, and above, the critical field strength, also known as the Schwinger limit~\cite{intro_1}. In case of an electric field:$\epsilon_{cr}=\frac{m_{e}^2c^3}{e\hbar}$ $\approx 1.32\times 10^{18} {V}/{m}$, where $m_{e}$ and $e$ are electron mass and charge respectively, $c$ is a speed of light and $\hbar$ is reduced Plank’s constant. This field accelerates an electron to the energy equivalent to its mass at a
distance of the Compton wavelength $\lambda$ = $\hbar$/mc, and leads to a possibility of spontaneous $e^{+}e^{-}$ pair generation.  These collisions will allow us to study the non-perturbative physics near or even above the Schwinger limit. The LUXE experiment will focus on the non-linear processes of Compton scattering, Breit-Wheeler pair production and trident pair production~\cite{bw_1}.
LUXE also offers new opportunities to directly search for new particles from physics beyond the Standard Model (BSM)~\cite{intro_2}.

\section{LUXE experimental set-up}
\begin{figure}[!htb]
\centering
 \includegraphics[height=5.17cm,keepaspectratio]{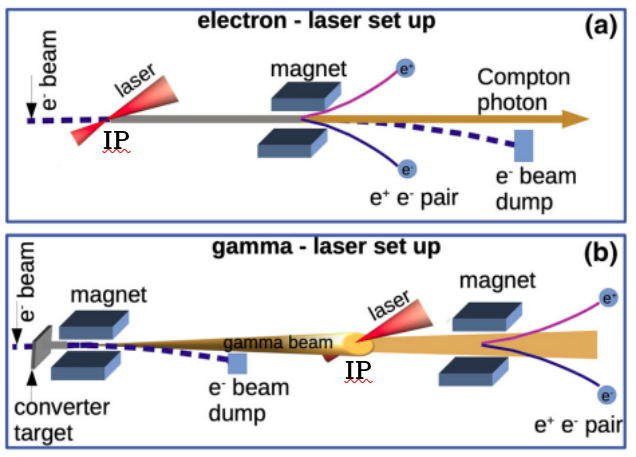}
  \caption{Sketch of the \textbf{(a)} e-laser and \textbf{(b)} $\gamma$-laser set-up.}\label{fig:luxe_a_b}
\end{figure}

Two phases are foreseen for the future LUXE experiment ~\cite{intro_3}. 
Schematic layouts of the experiment are shown in Figure~\ref{fig:luxe_a_b}  for the two configurations envisaged for the e-laser and the $\gamma$-laser. LUXE will measure electrons, positrons and photons and the setup of the experiment conceptually consists of electron and positron spectrometer, photon spectrometer and photon measuring subsystems.
 In the e-laser scenario the electron beam collides with the laser pulse at the interaction point (IP) located in the vacuum of the interaction chamber and the produced particles are confined to the narrow cone along the
primary beam and propagate towards the electron and positron spectrometer. For the $\gamma$-laser scenario the initial electron beam hits the target upstream of the IP. The spectrometer located after the target, equipped with a scintillator screen and Cherenkov gas detectors, will register electrons and positrons from conversions to estimate the number of produced photons.
\subsection{Optical laser}
The high power laser beam for LUXE has two options. In Phase 0, an optical ($\lambda$ = 800 $nm$) fs pulsed Ti-Sapphire laser system capable of providing pulse peak powers of up to 40 TW, will be used. In Phase 1, the beam will be upgraded to a commercial 350 TW laser. During the data-taking the dimensionless parameters ($\xi$ - the quantum parameter and $\chi$ - the intensity of the laser field) will be varied by de-focusing and re-focusing the laser beam at the IP. The crossing angle between the electron and the photon beam is expected to be about $18^\circ$. The laser pulse will be 30 fs long, and the repetition rate will be 1 Hz. This will leave 9 Hz of electron only data for background studies.

\subsection{Detectors}
One of the challenges of the LUXE experiment is to measure the energy spectra and fluxes of the $e^{-}$, $e^{+}$ and photons produced at the IP in both envisaged run-modes. The particle rates vary significantly. In some regions they are as low as 0.1 particles per bunch crossing, placing a high demand on the background rejection capabilities of the detector, and in other regions there are more than $10^{7}$ particles, placing high demand on the linearity and the radiation tolerance of
the device. In the low-rate regions, the detection relies on a silicon pixel tracker and a high-granularity compact calorimeter, while in the high-flux regions scintillation screens and Cherenkov detectors are the technologies selected.

\begin{figure}[!htb]
\centering
  \includegraphics[height=3.7cm,keepaspectratio]{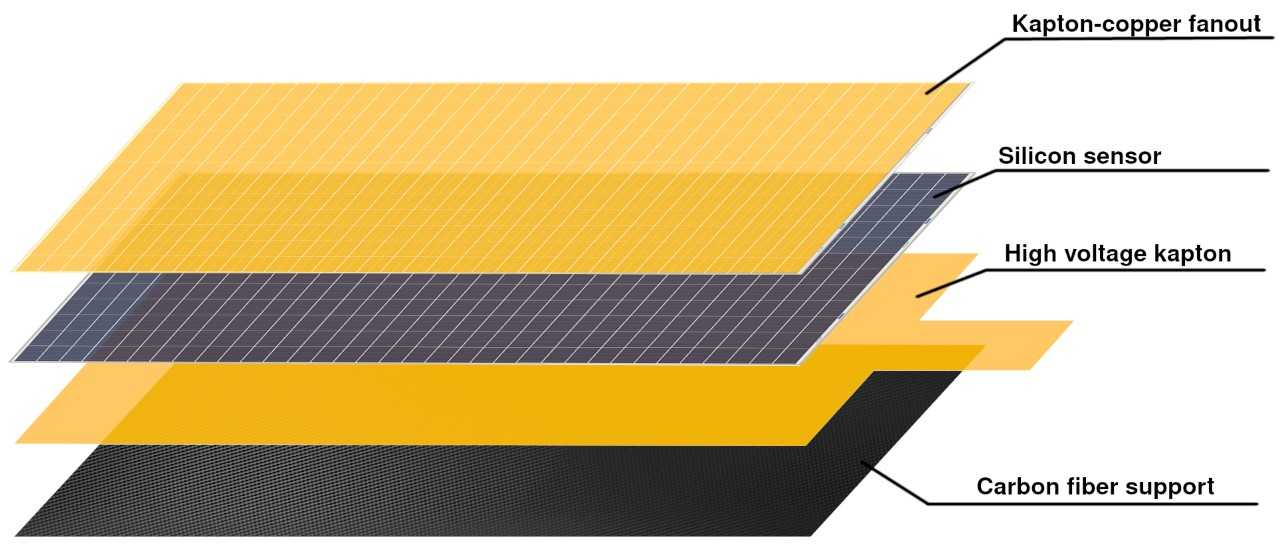}
  \caption{The structure of the assembled detector plane using a Kapton fan-out for the
signal routing.}\label{ecal_fig}
\end{figure}

\begin{figure}[!htb]
\centering
 \includegraphics[height=5.55cm,keepaspectratio]
 {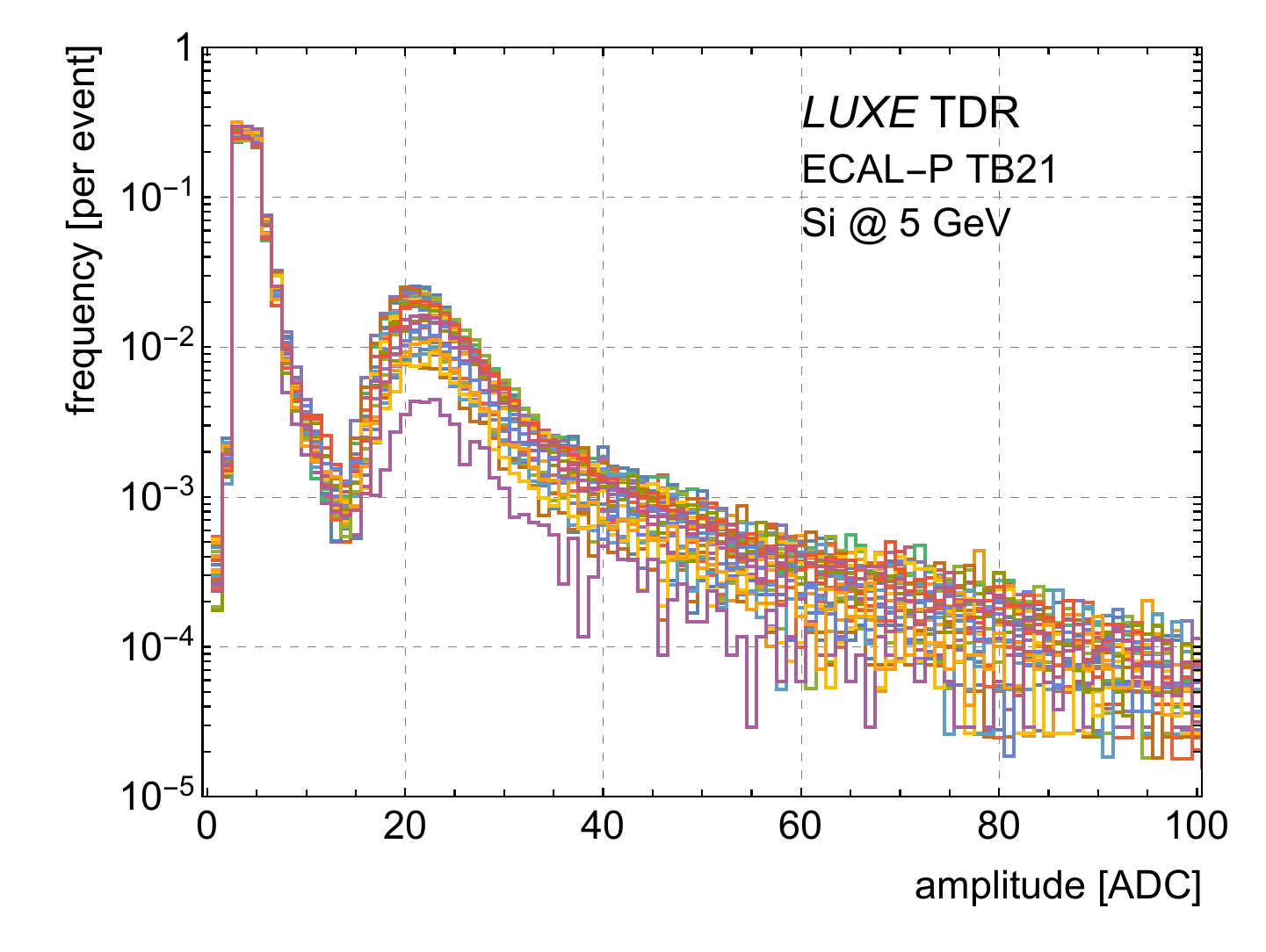}
  \caption{Signal distribution for all Si sensor channels.}\label{sig_gaas}
\end{figure}

Essential for the performance of the EM calorimetre (ECAL) are high granularity for very good position resolution, compactness, i.e.
a small Molière radius, to ensure a high spatial resolution of local energy deposits, and good energy resolution
to measure and infer the spectrum of positrons. To meet these requirements, technology developed by the FCAL collaboration is used~\cite{shower_2}. 
The ECAL is designed as a sampling calorimeter composed of 3.5 mm thick tungsten absorber plates, and assembled sensor planes placed in a 1 mm gap between absorber plates. A sketch of the structure of the assembled sensor plane using a Kapton fan-out is shown in Figure~\ref{ecal_fig}.
These sensors were measured in November 2021 at the DESY-II Synchrotron using electrons with energies between 1 and 5 GeV. The preliminary results of the signal distribution in a Si sensor is shown in Figure~\ref{sig_gaas}. As can be seen a clear signal of minimum ionising particles is visible, well separated from the noise. The size of the signal matches the expectations from the energy deposited by the relativistic electrons in the sensor material. 
The data analysis is in progress. These tests will be repeated in fall 2022 to perform a detailed scanning of the sensors uniformity, cross talk and edge effects using the telescope.
\section{Conclusions}
LUXE will be one of the first experiments to explore QED in the uncharted strong-field frontier. Major components developed by the FCAL Collaboration are considered for the LUXE experiment. The experimental setup that can be adapted to a large dynamic range has been designed and is going to be tested and installed in the XFEL.EU tunnel close to DESY - Osdorfer Born area. The first data taking is scheduled for 2025/2026.

\section*{Acknowledgments}
This work was partially supported by the Romanian Ministry of Research,
Innovation, and Digitisation, grant no. 16N/2019 within the National Nucleus Program. The measurements leading to these results have been performed at the Test Beam Facility at DESY Hamburg (Germany), a member of the Helmholtz Association (HGF).


\end{document}